\documentclass[twocolumn,preprintnumbers,amsmath,amssymb,superscriptaddress]{revtex4}
\usepackage{graphicx}
\usepackage{dcolumn}
\usepackage{bm}
\usepackage{soul}
\usepackage{color}
\usepackage{epstopdf}
\usepackage[version=3]{mhchem}
\usepackage{lipsum}
\usepackage[outercaption]{sidecap}
\usepackage{floatrow}
\usepackage{chngcntr}

\begin{document}

\preprint{APS/123-QED}

\title{Theory of the Spatial Transfer of Interface-Nucleated Changes of Dynamical Constraints and Its Consequences in Glass-Forming Films}
\author{Anh D. Phan}
\affiliation{Department of Physics, University of Illinois, 1110 West Green St, Urbana, Illinois 61801, USA}
\author{Kenneth S. Schweizer}
\affiliation{Department of Materials Science and Chemistry, Frederick Seitz Materials Research Lab, University of Illinois at Urbana-Champaign}
\email{kschweiz@illinois.edu}

\date{\today}

\begin{abstract}
We formulate a new theory for how caging constraints in glass-forming liquids at a surface or interface are modified and then spatially transferred, in a layer-by-layer bootstrapped manner, into the film interior in the context of the dynamic free energy concept of the Nonlinear Langevin Equation (NLE) theory approach. The dynamic free energy at any mean location (cage center) involves contributions from two adjacent layers where confining forces are not the same. At the most fundamental level of the theory, the caging component of the dynamic free energy varies essentially exponentially with distance from the interface, saturating deep enough into the film with a correlation length of modest size and weak sensitivity to thermodynamic state. This imparts a roughly exponential spatial variation of all the key features of the dynamic free energy required to compute gradients of dynamical quantities including the localization length, jump distance, cage barrier, collective elastic barrier and alpha relaxation time. The spatial gradients are entirely of dynamical, not structural nor thermodynamic, origin. The theory is implemented for the hard sphere fluid and diverse interfaces which can be a vapor, a rough pinned particle solid, a vibrating (softened) pinned particle solid, or a smooth hard wall. Their basic description at the level of the spatially-heterogeneous dynamic free energy is identical, with the crucial difference arising from the first layer where dynamical constraints can be weaken, softened, or hardly changed depending on the specific interface. Numerical calculations establish the spatial dependence and fluid volume fraction sensitivity of the key dynamical property gradients for five different model interfaces. Comparison of the theoretical predictions for the dynamic localization length and glassy modulus with simulations and experiments for systems with a vapor interface reveal good agreement. The present advance sets the stage for using the Elastically Collective NLE theory to make quantitative predictions for the alpha relaxation time gradient, decoupling phenomena, $T_g$ gradient, and many film-averaged properties of both model and experimental (colloids, molecules, polymers) systems with diverse interfaces and chemical makeup.
\end{abstract}

\maketitle

\section{Introduction}
Activated dynamics, mechanical properties and vitrification in thin films of glass-forming liquids of diverse chemical nature (atoms, colloids, molecules, polymers) with highly varied boundary conditions is a problem of great intrinsic scientific interest, which additionally may (or may not) shed light on the physics of the bulk glass transition \cite{1,2,3,4,5}. Thin films are also important in many materials applications \cite{6,7,8,9}. Despite intense experimental, simulation and theoretical effort over the past two decades \cite{2,5,10,11,12,13,14,15,16,17,18}, the key physical mechanisms underlying the observed phenomena remain not very well understood. We believe that this reflects the complexity of activated relaxation in bulk liquids \cite{10} in concert with the formidable complications of geometric confinement, interfaces and spatial inhomogeneity.

A particularly rich aspect of thin films is the qualitatively varied impact of boundary conditions. Free standing thin films with two vapor interfaces, or semi-infinite thick films with one vapor interface/surface, are the simplest realizations of confined systems. Extensive experimental \cite{1,2,3,19,20,21,22,23,24,25,26,27,28} and simulation \cite{2,5,14,29,30,31,32}  efforts suggest a spatially inhomogeneous large speed up of structural relaxation with mobile layers extending rather deep into the film with correspondingly large film-averaged reductions of the glass transition temperature, $T_g$. In contrast, experiments and simulations find that near a solid substrate the dynamics is very non-universal -- it can modestly speed up, slow down drastically, or hardly change at all relative to the bulk \cite{2,5,13,14,24,25,33,34,35,36}. The origin of such complexity often seems puzzling. A confining surface or substrate can be topographically smooth or rough, can promote liquid adsorption or not, and can have a mechanical stiffness varying from infinitely rigid (pinned particles) to a soft surface \cite{37,38,39} to even liquid substrates \cite{40} that are thermodynamically hard but dynamically fluid. It appears all these features are important, often qualitatively, for determining the glassy dynamics of real world films.

Recently, a quantitative force-level statistical mechanical approach for structural (alpha) relaxation in isotropic colloidal, molecular and polymer bulk liquids, the "Elastically Collective Nonlinear Langevin Equation" (ECNLE) theory \cite{41,42,43,44,45,46}, has been developed and generalized to treat glassy dynamics in free-standing films \cite{47,48,49}. Structural relaxation is described as a coupled activated process involving a large amplitude cage-scale particle hopping event that is facilitated by a small amplitude longer-range collective elastic deformation of the surrounding liquid. Quantitative tractability for molecular and polymeric liquids is achieved based on an a priori mapping of chemical complexity to a thermodynamic-state-dependent effective hard sphere fluid \cite{42,45}. The theory for free-standing films predicts strongly accelerated and spatially inhomogeneous relaxation for purely dynamical reasons. 

Most recently, Phan and Schweizer \cite{50} formulated an improved technical treatment of the collective elasticity aspect in free-standing thin films and semi-infinite thick films with vapor interfaces, and addressed qualitatively new questions. For example, the mobile layer length scale is predicted to grow strongly with cooling, and correlates nearly linearly with the dynamic barrier deduced from the bulk alpha time. A new type of spatially inhomogeneous "decoupling" was predicted, an effect first discovered by Simmons and coworkers using computer simulation in the weakly supercooled regime \cite{57}. Specifically, this type of decoupling corresponds to a remarkable effective factorization of the total barrier into its bulk temperature-dependent value multiplied by a function that depends only on location in the film. Quantitative no-fit-parameter comparisons of the theory for free standing films with experiment and simulation for $T_g$ shifts of polystyrene and polycarbonate are in reasonable accord with the theory, and testable predictions were made \cite{19,51,52,53,54}

However, major puzzles remain even for films with vapor interfaces. Conceptual ones include precisely how mobility changes are nucleated at an interface or surface, and how they are "propagated" or transferred deep into the film. How such questions can be theoretically addressed for films with solid interfaces is open. Crucial motivations for the present article are puzzles such as the long standing simulation finding that the relaxation time gradient for free standing and solid substrate films appears to have (to leading order) a "double exponential" form \cite{13,14,31,55,56,57,58}. This behavior implies the effective barrier varies roughly in an exponential manner with distance from an interface. However, the associated length scale only modestly grows with cooling, and appears to already saturate in the lightly supercooled regime probed in simulation \cite{55,56,57,58}. Such behavior is in apparent disagreement with entropy crisis or thermodynamic-based theories of glassy dynamics which argue the relevant length scale should continue to grow all the way down to the laboratory vitrification temperature \cite{10,16,56}. A seemingly related behavior revealed by simulation is a spatial dependence of the "decoupling exponent" that varies roughly exponentially with distance from the interface, and the strong correlation of this behavior with gradients of the activation barrier \cite{57}. 

The present paper reports the first and most critical advance required to generically address the above issues within the ECNLE theoretical framework. Specifically, we formulate a new treatment of how local dynamical constraints, quantified via a cage scale "dynamic free energy", are modified at an interface, and how they are transferred into the film. The ideas are applied to study the spatial dependence of the particle localization length and glassy elastic modulus, and also to establish how all dynamic free energy properties that determine the total activation barrier and alpha time gradient are modified. This sets the stage for future efforts that will employ ECNLE theory to quantitatively predict the alpha relaxation time gradient and other properties for films with diverse boundary conditions.

The remainder of the article is as follows. We briefly review in Section II the key elements of the existing ECNLE theory of bulk liquids and vapor interface films. Section III presents our new formulation of how cage scale dynamical constraints are modified for various soft and hard interfaces. Five different hard, soft and vapor interfacial models are considered. Application to treat the dynamic localization length and glassy modulus in films is the subject of Section IV, and quantitative no-fit-parameter comparisons are made with experiment and simulation. Section V establishes how all other features of the dynamic free energy in films are modified. The paper concludes with a discussion in Section VI. The Appendix compares predictions for the localization length obtained from two different formulations of the new theoretical idea.  

\section{Background: ECNLE Theory of Bulk Liquids and Free-Standing Thin Films}
For context, we briefly review the present state of ECNLE theory for bulk liquids \cite{41,42,43,44,45,46} and free standing thin films \cite{47,48,49,50} in the simplest context of spherical particle liquids; all details are in prior papers. In this article, we will implement the new ideas for the foundational hard sphere system.
\subsection{Bulk Liquids}
Consider a one-component liquid of spherical particles (diameter, $d$) of packing fraction $\Phi$. The fundamental theoretical quantity is an angularly-averaged particle displacement-dependent "dynamic free energy", $F_{dyn}(r)=F_{ideal}(r)+F_{caging}(r)$ , the derivative of which is the effective force on a moving particle in a stochastic nonlinear Langevin equation (NLE) \cite{59}:
\begin{eqnarray} 
\frac{F_{dyn}(r)}{k_BT} &=& -3\ln r-\rho\int\frac{d\mathbf{q}}{(2\pi)^3}\frac{S(q)C^2(q)}{1+S^{-1}(q)}\nonumber\\
&\times&\exp\left[-\frac{q^2r^2}{6}\left(1+S^{-1}(q) \right) \right] \nonumber\\
&=& \frac{F_{ideal}(r)}{k_BT}+\frac{F_{caging}(r)}{k_BT},
\label{eq:1}
\end{eqnarray}
where $\beta = (k_BT)^{-1}$, $k_B$ is the Boltzmann's constant, $T$ is temperature, $\rho$ is number density, $r$ is the displacement of a particle from its initial position, $S(q)$ is the structure factor, $q$ is wavevector, and $C(q)=\rho^{-1}\left[1-S^{-1}(q) \right]$ is the direct correlation function. The leading term in Eq.(\ref{eq:1}) is an ideal entropy-like contribution that favors the fluid state, and the second term is due to interparticle forces which favors cage localization. The latter is determined from knowledge of fluid density and pair liquid structure. As the density (or temperature) exceeds (goes below) a critical value, a local barrier $F_B$ in $F_{dyn}(r)$ emerges (at $\Phi \approx 0.43$  for hard spheres \cite{59} based on Percus-Yevick theory \cite{60} input) signaling transient localization. Figure 1 shows an example dynamic free energy, its ideal and caging components, and defines key length and energy scales including the localization length, $r_L$, barrier location, $r_B$, jump distance, $\Delta r = r_B - r_L$, and local cage barrier, $F_B$.

\begin{figure}[htp]
\includegraphics[width=8cm]{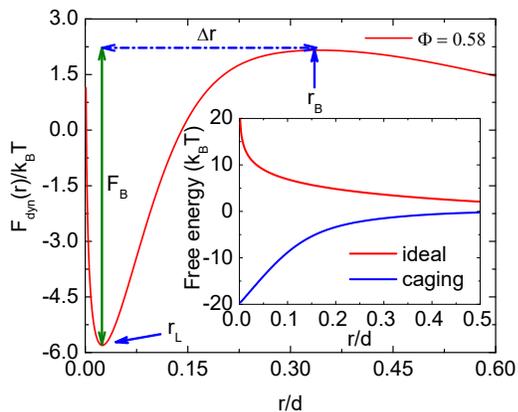}
\caption{\label{fig:1}(Color online) Dynamic free energy as a function of reduced particle displacement for a hard sphere fluid of packing fraction $\Phi = 0.58$ ; important length and energy scales are defined. The inset shows the corresponding ideal and caging components of $F_{dyn}(r)$.}
\end{figure}

For hard sphere fluids with barriers beyond a few $k_BT$, much insight has been gained based on the so-called “ultra-local” analytic analysis \cite{61}. The crucial result is that, to leading order, all aspects of the dynamic free energy enter via a universal function multiplied by a single "coupling constant", $\lambda$ \cite{61}:
\begin{eqnarray} 
F_{caging}(r) = \lambda(\Phi).f_{cage}(r/d), \quad \lambda(\Phi) \propto \Phi g(\Phi)^2,
\label{eq:2}
\end{eqnarray}
where $g(d)$ is the contact value of $g(r)$. The first equality in Eq.(\ref{eq:2}) is a factorization-like property which implies the functional form of the caging dynamic free energy (and corresponding force, ($-\partial F_{caging}(r)/\partial r$) is, to leading order, not dependent on thermodynamic state. The local structure and packing fraction enter solely in a multiplicative manner via a coupling constant, $\lambda$. This is a striking prediction of NLE theory that holds when barriers are relatively high and motion is strongly activated. Detailed analysis shows the coupling constant can be physically interpreted as proportional to an effective mean square caging force experienced by a tagged particle. It is dominated by nearest neighbor forces for short range interactions (high $q$ contributions dominate in Eq.(\ref{eq:1})). Prior analytic analysis has derived \cite{61}: 
\begin{eqnarray} 
r_L &\approx& \frac{\sqrt{3\pi}}{4\Phi g^2(d) } \propto \lambda^{-1} \propto (\beta F_B)^{-1}, \nonumber\\
r_B &=& \frac{1}{q_c}\sqrt{3\ln(4\Phi g^2(d))}, \quad q_c = \pi/d.
\label{eq:3}
\end{eqnarray}
The predicted relation $d/r_L \propto \beta F_B \propto \lambda$ connects short and long time dynamics, a hallmark of NLE theory. The dynamic (relaxed high frequency) shear modulus, $G'$, is predicted (not assumed) to obey a micro-rheology like relation \cite{41,61}:
\begin{eqnarray} 
G'\approx \frac{9\Phi k_BT}{5\pi dr_L^2}.
\label{eq:4}
\end{eqnarray}
These connections remain useful for thermal liquids since they are a priori mapped to effective hard sphere fluids \cite{42,45}. In Eq.(\ref{eq:2}), the coupling constant then becomes a function of temperature, pressure and chemistry. The connections also remain useful in thin films. 

\begin{figure}[htp]
\includegraphics[width=8cm]{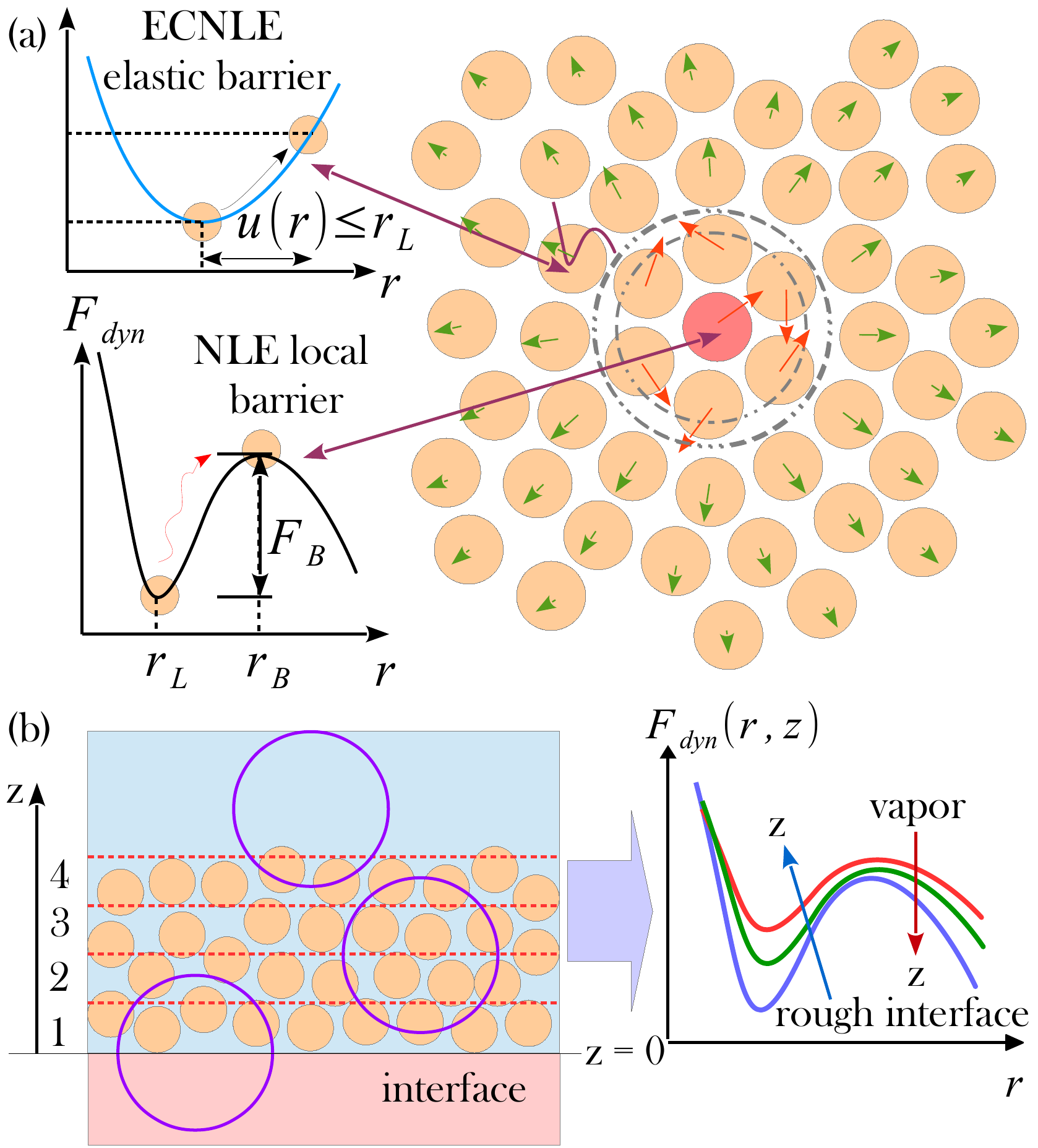}
\caption{\label{fig:2}(Color online) (a) Schematic of the fundamental relaxation event in bulk liquid ECNLE theory involving two coupled physical processes: (1) local/cage-scale hopping as described by the dynamic free energy, and (2) a nonlocal/spatially longer range collective harmonic elastic motion outside the cage region required to allow the large amplitude local rearrangement to occur. Various key length and energy scales are indicated. (b) Cartoon illustration of the layer-like model of the surface nucleated dynamic caging constraint transfer idea and spatial variation of the dynamic free energy.}
\end{figure}

In ECNLE theory, large amplitude hopping is strongly coupled to a long range collective elastic spontaneous fluctuation of all particles outside the cage required to create the small amount of extra space to accommodate a hop; conceptual elements are sketched in Fig. \ref{fig:2}a. The radially-symmetric solution for the required elastic displacement field decays as an inverse square power law \cite{41,62}:
\begin{eqnarray}
u(r)&=&\Delta r_{eff}\frac{r_{cage}^2}{r^2}, \qquad r\geq r_{cage} \nonumber\\
\Delta r_{eff} &\approx& 3\Delta r^2/32r_{cage} \leq r_L.
\label{eq:5}
\end{eqnarray}
The amplitude is set by a small mean cage expansion length, $\Delta r_{eff}$ , which follows from assuming each particle in the cage independently hops in a random direction by $\Delta r$. The elastic barrier is determined by summing over all harmonic displacements outside the cage region thereby yielding \cite{41}

\begin{eqnarray}
F_{elastic} &=& \rho\frac{K_0}{2}\int_{r_{cage}}^\infty dr 4\pi r^2 u^2(r)g(r)\nonumber\\
&\approx& 12K_0\Phi\Delta r_{eff}^2\left(\frac{r_{cage}}{d}\right)^3,
\label{eq:6}
\end{eqnarray}
where $r$ is relative to the cage center, and $K_0=3k_BT/r_L^2$ is the curvature of the dynamic free energy at its minimum. The sum of the coupled (and in general temperature and density dependent) local and elastic collective barriers determine the total barrier for the alpha process, $F_{total} = F_B + F_{elastic}$. A generic measure of the structural or alpha relaxation time follows from a Kramers calculation of the mean first passage time for barrier crossing. For barriers in excess of a few $k_BT$ one has \cite{41,59}:
\begin{eqnarray}
\frac{\tau_\alpha}{\tau_s} = 1+ \frac{2\pi(k_BT/d^2)}{\sqrt{K_0K_B}}\exp{\frac{F_B+F_{elastic}}{k_BT}},
\label{eq:7}
\end{eqnarray}
where $K_B$ is the absolute magnitude of the barrier curvature. The alpha time is expressed in units of a "short time/length scale" relaxation process (cage-renormalized Enskog theory), $\tau_s$, the explicit formula for which is given elsewhere \cite{41,42} . Physically, it captures the alpha process in the absence of strong caging defined by the parameter regime where no barrier is predicted (e.g., $\Phi < 0.43$ for hard spheres). The latter condition corresponds to being below the naïve mode coupling theory (NMCT \cite{59,63}) ideal dynamic glass transition which in ECNLE theory is manifested as a smooth crossover.

The theory can be applied to any spherical particle of colloidal fluid, and to molecular and polymeric liquids based on an appropriate mapping \cite{42,45}; here we consider only the hard sphere fluid. To place our calculations in broader context, we recall how packing fraction, reduced temperature and alpha time are related for the prototypical glass-forming molecular liquid orthoterphenyl (OTP) \cite{42}: $\Phi = 0.53, 0.55, 0.57, 0.59, 0.61$, corresponds to $T/T_g \approx 1.53, 1.40, 1.27, 1.15, 1.04$, $\tau_\alpha \sim$ 1.64 ps, 6.5 ps, 183 ps, 122 ns, 0.061 s,where $\tau_\alpha = 100$ s at $T=T_g$.
\subsection{Vapor Interface Films}
For films with interfaces every property (thermodynamic, structural, dynamic) is spatially heterogeneous and anisotropic. Treating such complexity theoretically is intractable. In the past a minimalist approach was adopted based on the hypothesis that the most important effects are purely dynamical with no changes of thermodynamics or structure in the film \cite{47,48,49,50}. This idea is consistent with recent machine-learning based analysis of simulations of free standing films which found the large dynamical changes are not related to any change of structural or static properties \cite{64}. It also is relevant for simulation studies of films performed under so-called "neutral confinement" conditions \cite{13,14,55,56,65} where the solid substrate is constructed to have no effect on liquid packing. 

Of course, in real thin films there are changes of thermodynamic properties, the one-body density profile perpendicular to the interface, and (anisotropic) packing correlations near an interface. But these changes are usually small, highly localized near the interface in a dense liquid, depend sensitively on the nature of the interface or surface, and are chemically specific. In this article, for both simplicity and our desire to focus on the purely dynamical physics, we ignore such complications and adopt a step-function density profile in the direction orthogonal to the interface and a density or volume fraction identical to that in the bulk liquid.

How a vapor interface modifies the alpha process in the prior ECNLE theory involves two coupled effects: (i) local caging as encoded in the dynamic free energy, and (ii) the collective elastic displacement field and associated barrier. The cage remains the elementary dynamical unit and is characterized locally by (pre-averaged) isotropic symmetry. The goal is to predict how it changes as a function of distance from the film interface, $z$. A zeroth order approach \cite{47,48,49,50} [47-50] for free standing films was constructed as follows. For point (i), near the surface ($0 \leq z \leq r_{cage}$ where for a sharp interface the center of particles of the first layer define $z = 0$) caging constraints are softened due to losing nearest neighbors. The fraction of bulk cage particles present at location $z$ follows from geometry as \cite{47}:
\begin{eqnarray} 
\gamma(z) = \frac{1}{2} - \left(\frac{z}{r_{cage}}\right)^3\left[\frac{1}{4}-\frac{3}{4}\left(\frac{r_{cage}}{z}\right)^2 \right].
\label{eq:8}
\end{eqnarray}
For $z = 0$, $\gamma(z) = 0.5$ corresponds to losing one half of the nearest neighbors. It can also be thought of as setting to zero the collective dynamic Debye-Waller factor $e^{-q^2r^2/6S(q)}$ in Eq.(\ref{eq:1}) for the fraction $1-\gamma(z)$ of particles missing from the effective cage, while the remaining particles (quantified by the factor $\gamma(z)$) have the same collective dynamic Debye-Waller factor as in the bulk. For $z = r_{cage}$, the full cage is recovered and $\gamma(z) = 1$. This is a highly local approximation, where surface-induced mobility is assumed to not extend into the film beyond the cage radius (it is this approximation that is re-visited in the present work). The dynamic free energy is thus modified as \cite{47}:  
\begin{eqnarray} 
F_{dyn}(r) = -3k_BT\ln r + \gamma(z)F_{caging}(r).
\label{eq:9}
\end{eqnarray}

Near the surface all properties of the dynamic free energy behave as a liquid with weaker dynamical constraints. Importantly, note the multiplicative manner the interface modifies dynamical constraints where $F_{caging}(r)$ remains the same as in the bulk. Given Eq.(\ref{eq:2}), this implies, to leading order, a "double factorization" type of mathematical structure: $F_{caging}(r) \gamma(z).\lambda(\Phi).f_{cage}(r/d)$. This can have potentially profound consequences. For example, to leading order a type of "corresponding states" behavior is expected for the caging constraints since $f_{cage}(r/d)$ is universal but a continuum of values of $(z,\Phi)$ in principle exist such that the net amplitude of the caging dynamic free energy and force (determined by the product $\gamma(z).\lambda(\Phi)$ remains constant. To address point (ii) above, a simple "cutoff" of the bulk isotropic elastic field assumption was adopted, formulated in two technically different, but qualitatively the same, manners [41]. Since this article focuses solely on point (i), we do not elaborate further, except to emphasize that all the information required to determine the elastic barrier (jump distance and dynamic localization length since $K_0 = 3k_BT/r_L^2$) also follows from knowledge of the dynamic free energy in the film.

\section{New Formulation of Interface-Induced Spatially Inhomogeneous Caging Constraints}
The first and foremost critical issue is: (i) how is the caging force modified near an interface? Prior work \cite{38,39,40,41} for a vapor surface assumed that beyond a radius $r_{cage}\sim 1.3-1.5d$ the dynamic free energy recovers its bulk form. Additional simplifications were invoked to render the theory tractable and/or for internal consistency with the bulk formulation. (ii) The liquid-vapor interface is perfectly sharp. (iii) The ensemble-averaged pair structure, liquid density and thermodynamic properties are unchanged in the film. (iv) The mobility of all particles in a spherical cage region are the same. Assumptions (ii) and (iii) can be relaxed at the expense of technical complexity. Assumption (iv) pre-averages dynamic heterogeneity inside the cage scale of $\sim 3d$, retaining the spirit of bulk NLE theory. 

Here we propose a new general formulation of the dynamic free energy idea for films that we believe qualitatively improves the treatment of (i) and (iv). Point (i) is the most fundamental, and we aim to understand how mobility near the surface can affect particles in a layer directly above it, and how such a gradient of dynamical constraints extends further into the film. For a vapor interface where dynamics speeds up, one could view this as a form of "dynamic facilitation", albeit of literal broken spatial symmetry origin of different physical origin than in an isotropic bulk fluid. \cite{10} For a solid surface that slows down particles near it, the effect would be akin to "anti-facilitation".

We first recall that bulk NLE theory is built on the single particle (naive) version of ideal mode coupling theory (so-called NMCT \cite{59,63}) as encoded in a self-consistent nonlinear equation for strict kinetic arrest based on an ensemble-averaged localization length. NMCT relates pair structure, forces, thermodynamic state and caging constraints. Given the film problem is more complex, we first explore two different approaches which are in the same spirit physically. Both adopt a finer resolution of space than a cage diameter to formulate dynamic constraints, namely a "layer" which can be interpreted as a region of one particle diameter or cage radius thickness; here we adopt the former perspective. See Fig. \ref{fig:2}b for a sketch. The layer picture is a conceptual device to quantify constraints in a spatially discrete manner. It does not require any density gradient perpendicular to the flat interface.

The first approach is in the NMCT framework and only addresses the ideal glass question. The second general approach is formulated directly in terms of the dynamic free energy concept. As shown in the Appendix, for the only question these two formulations can both address, the gradient $r_L(z)$, the numerical results are similar. The second approach is the focus of our present and future efforts.
\subsection{NMCT Gaussian Dynamical Formulation}
The NMCT self-consistent localization relation for an ideal glass in the isotropic bulk is \cite{59}: 
\begin{eqnarray} 
\frac{9}{r_L^2} = \int \frac{d\mathbf{k}}{(2\pi)^3}|kC(k)|^2\rho S(k)e^{-k^2r_L^2(1+S^{-1}(k))/6},
\label{eq:10}
\end{eqnarray}
where $\left<r^2(t\rightarrow \infty) \right> \equiv r_L^2$, and $e^{-k^2r_L^2/6}$ and $e^{-k^2r_L^2/6S(k)}$ are the kinetically arrested single and collective dynamic propagators (Debye-Waller factors), respectively. Per Fig. \ref{fig:2}b, for a film we change perspective to a finer resolution of the cage corresponding to a layer-like model or (in practice) resolving a cage into two halves. Since in-plane particle localization is taken to be uniform at a given distance from the interface, the arrested dynamical state in layer $i$ (or $z = (i-1)d$ in terms of spatial position) is described by $r_{L,i}$. We continue to adopt the physical picture of a cage of diameter ~3d surrounding a tagged particle which encapsulates particles from three layers. Focusing on a particle at the cage center, we view it as experiencing forces from an equal number of particles above and below (if present) it. As our starting point ansatz, a central particle is modeled as experiencing two types of dynamical environments in a film depending on its distance from the interface. Within each half of a cage, we average over particle mobility, in contrast to bulk NLE theory which averages over all particles in a spherical cage. Now, based on the idea that dynamical inhomogeneity is initiated at the interface, the caging constraints on a particle in a given layer are constructed in a democratic fashion. This corresponds to a collective Debye-Waller factor in Eq. (\ref{eq:10}) that has two contributions yielding a modified self-consistent NMCT equation: 

\begin{eqnarray} 
\frac{9}{r_{L,i}^2} &=& \int \frac{d\mathbf{k}}{(2\pi)^3}|kC(k)|^2\rho S(k)e^{-k^2r_L^2/6} \nonumber\\
&\times& \left[\frac{1}{2}e^{-k^2r_{L,i}^2/6S(k)} + \frac{1}{2}e^{-k^2r_{L,i-1}^2/6S(k)} \right].
\label{eq:11}
\end{eqnarray}
The first (second) term inside the bracket corresponds to the dynamic constraints from half of a particle cage of center assigned to layer $i (i-1)$.

We consider a thick film with one vapor or pinned solid interface. For the former, tagged particles in the first layer do not experience forces from the underlayer since there are no particles. Thus, for the first layer one has from Eq.(\ref{eq:11}) a closed equation:
\begin{eqnarray} 
\frac{9}{r_{L,1}^2} = \int \frac{d\mathbf{k}}{(2\pi)^3}|kC(k)|^2\rho S(k)e^{-k^2r_L^2/6} \left[\frac{1}{2}e^{-k^2r_{L,1}^2/6S(k)} \right].
\label{eq:12}
\end{eqnarray}
This is identical to the Mirigian-Schweizer (MS) approach \cite{47,48,49} for $z=0$. For a supported film, we first consider the case where the substrate is modeled as a quenched fluid composed of literally pinned particles of the same size, density and pair structure as the mobile particle liquid that defines the film (often called "neutral confinement"). Then the first layer localization length is determined by Eq.(\ref{eq:11}) with $r_{L,0}^2 = 0$. The localization length of particles in first layer of the mobile liquid is thus:
\begin{eqnarray} 
\frac{9}{r_{L,1}^2} &=& \int \frac{d\mathbf{k}}{(2\pi)^3}|kC(k)|^2\rho S(k)e^{-k^2r_L^2/6} \nonumber\\
&\times& \left[\frac{1}{2}e^{-k^2r_{L,1}^2/6S(k)}+ \frac{1}{2} \right].
\label{eq:13}
\end{eqnarray}
Importantly, an equation identical to that above follows if we employ our recent theory \cite{66} of the bulk pinned-mobile hard sphere system with the fraction of pinned particles set to 0.5. The reason is that in both cases a tagged particle experiences one half of its constraints from immobile but otherwise identical particles. This exposes a key assumption: in broken symmetry films it is the number of particles that are mobile versus immobile in a spherically-averaged cage which quantifies (to leading order) the dynamical constraints on a tagged particle in a cage; the precise spatial arrangement is angularly pre-averaged. This essential approximation is what renders the theory tractable, and allows us to think and calculate in a manner analogous to prior NLE theory work in bulk and thin films. 

The full dynamic localization length gradient then follows immediately from the above ideas and Eq.(\ref{eq:11}). Note that the localization length in layer $i$ follows from knowledge of its analog in the underlayer $i-1$. Thus, one can predict the full gradient in a sequential layer-by-layer or bootstrapped manner starting at the surface, resulting in a simple physical picture and easy numerical solution.

\subsection{Dynamic Free Energy Formulation}
We now consider the problem directly from viewpoint of the dynamic free energy. The physical idea for introducing sub-cage resolution of dynamical constraints remains the same as above. Consider a particle at a cage center. We again assume dynamic constraints on it arise from equal contributions of particles in two adjacent layers. Given we assume packing structure is not changed in the film, the dynamic free energy in layer $i$ is:
\begin{eqnarray} 
F_{dyn}^{(i)}(r) = \frac{1}{2}F_{dyn}^{bulk}(r) + \frac{1}{2}F_{dyn}^{(i-1)}(r), \quad i \geq 1
\label{eq:14}
\end{eqnarray}
where $i=0$ is the first layer of the substrate. The "1/2-1/2" weighting form is the same as in Eq. (\ref{eq:11}). For the purpose of analyzing layer $i$, the constraints from the upper half of the cage are quantified as in the bulk. This is another key approximation, but one we believe is consistent with the assumed invariance of equilibrium structure in the film. But the particles in the lower half of a cage are affected by the interface in a manner that depends on the nature of, and distance from, the interface. Thus, the idea is again that film perturbations are nucleated in the first layer, and via modification of the caging part of the dynamic free energy are spatially transferred into the film. For the first liquid layer one has,
\begin{eqnarray} 
F_{dyn}^{(1)}(r) &=& \frac{1}{2}F_{dyn}^{bulk}(r) + \frac{1}{2}F_{dyn}^{(0)}(r) \nonumber\\
&=& F_{ideal}(r) + \frac{1}{2}F_{caging}^{bulk}(r) + \frac{1}{2}F_{caging}^{surface}(r), \nonumber\\
\label{eq:15}
\end{eqnarray}
where the crucial quantity is the "surface layer caging dynamic free energy", the last term above. The dynamic free energy of the film is constructed by iterating Eq.(\ref{eq:15}). For the second layer and third layers one has
\begin{eqnarray} 
F_{dyn}^{(2)}(r) &=& F_{ideal}(r) + \frac{1}{2}F_{caging}^{bulk}(r) + \frac{1}{2}F_{caging}^{(1)}(r), \nonumber\\
&=& F_{ideal}(r) + \left(\frac{1}{2}+\frac{1}{2^2}\right)F_{caging}^{bulk}(r) + \frac{1}{2}F_{caging}^{surface}(r), \nonumber\\
\label{eq:16}
\end{eqnarray}

\begin{eqnarray}
F_{dyn}^{(3)}(r) &=& \frac{1}{2}F_{dyn}^{bulk}(r) + \frac{1}{2}F_{dyn}^{(2)}(r) \nonumber\\
&=& F_{ideal}(r) + \left(\frac{1}{2}+\frac{1}{2^2}+\frac{1}{2^3}\right)F_{caging}^{bulk}(r)  \nonumber\\
&+& \frac{1}{2}F_{caging}^{surface}(r), 
\label{eq:17}
\end{eqnarray}
One can obviously write a general expression for the dynamic free energy in $n^{th}$  layer
\begin{eqnarray}
F_{dyn}^{(n)}(r) &=& F_{ideal}(r) + \left(1-\frac{1}{2^n}\right)F_{caging}^{bulk}(r)  +
 \frac{F_{caging}^{surface}(r)}{2^n}, \nonumber\\
 &=& F_{dyn}^{bulk}(r) + 2^{-n}\Delta F_{caging}(r)
\label{eq:18}
\end{eqnarray}
where
\begin{eqnarray}
\Delta F_{caging}(r) = F_{caging}^{surface}(r) - F_{caging}^{bulk}(r).
\label{eq:19}
\end{eqnarray}

The physical essence of this approach is effectively a hypothesis of a geometric-like transfer of dynamical constraint information nucleated at the surface into the film. The amplitude of the change of constraints enter via a difference in caging dynamic free energy (Eq.(\ref{eq:19})) which is expected to be positive (negative) for a pinned solid (vapor) surface. The generic form above implies the dynamic free energy varies essentially exponentially in space if one mathematically passes from a discrete layer description to a continuous space description:
\begin{eqnarray}
2^{-n} = e^{-n\ln2}=e^{-z/\xi}, \ce{where} \quad z = nd, \xi = d/\ln2.
\label{eq:20}
\end{eqnarray}
Importantly, the corresponding "decay length" is a universal constant of $\sim 1.4d$, but only at the most fundamental level of the caging dynamic free energy. Of course the latter is a theoretical construct that is not directly observable, and thus this simplicity does not generically apply for various dynamical properties derived from the dynamic free energy and full ECNLE theory. The amplitude of the change of dynamical constraints in Eq. (\ref{eq:19}) depends on chemistry, thermodynamic state, and nature of the surface. Moreover, the amplitude and $z$-dependence of caging constraints effectively factorize. Given the ultra-local analytic understanding of bulk NLE theory \cite{61} reviewed above, qualitatively one then expects the local barrier and all other key aspects of the dynamic free energy vary roughly exponentially as a function of distance from the interface (as shown numerically below). If true, this immediately provides a generic physical mechanism for the simulation observations of a "double exponential" form of alpha time gradients \cite{13,14,55,56,57,58}. 

Recall from the discussion below Eq. (\ref{eq:9}) of Section IIB that the fundamental form of the caging part of the dynamic free energy of the prior ECNLE theory \cite{50} for free-standing films obeyed the "double factorization" form.  Eqs. (\ref{eq:18})-(\ref{eq:20}) continue to obey this general form for the difference between the caging component of the dynamics in the bulk and at a location $z$ in the film. This property of the theory is expected to have many consequences. For example, as shown below, the spatial gradients of dimensionless ratios of a dynamic property in the film relative to in the bulk can often be (to leading order) invariant to temperature, volume fraction and chemistry. Moreover, the "corresponding states" structure mentioned in Section IIB continues to hold to leading order. 

We note that the existence of the simplicities described above rely on several physical ansatzes of the theory: high wavenumber dominance of the caging dynamic free energy, no changes of equilibrium properties in the film relative to the bulk, and the multiplicative manner that the location in the film variable modifies the dynamic free energy corresponding to a $z$-dependence that does not directly depend on thermodynamic state or chemistry. 

Finally, we comment on two fundamental theoretical aspects of our present formulation. First, Eq.(\ref{eq:14}) plus Eq.(\ref{eq:20}) may perhaps be interpreted as a forward-difference approximation of a gradient expansion of the dynamic free energy. This was not our perspective in formulating the theory, especially since dynamical spatial gradients are often very sharp for supercooled liquids near interfaces. Rather, we have chosen to formulate the theory in a discrete manner that explicitly acknowledges the finite size of particles and cages which are the elementary scales of NLE theory and the dynamic free energy concept. A second question is whether there could, or should, be an explicit coupling of layer $i$ with both layers $i-1$ and $i+1$, perhaps in the spirit of a $1-d$ Ising model. We note that such a formulation would introduce much additional technical and conceptual complexity since all layers become effective coupled and the dynamics of the entire film would need to be treated self-consistently. This is in contrast to our simpler formulation which has a layer-by-layer "bootstrapping" character. Moreover, our approach is in the spirit of the often invoked physical notion that dynamic changes at an interface "propagate" or are transferred in a directional manner from the interface into the film. 
\subsection{Specialization to a Specific Interface}
\begin{figure}[htp]
\includegraphics[width=8cm]{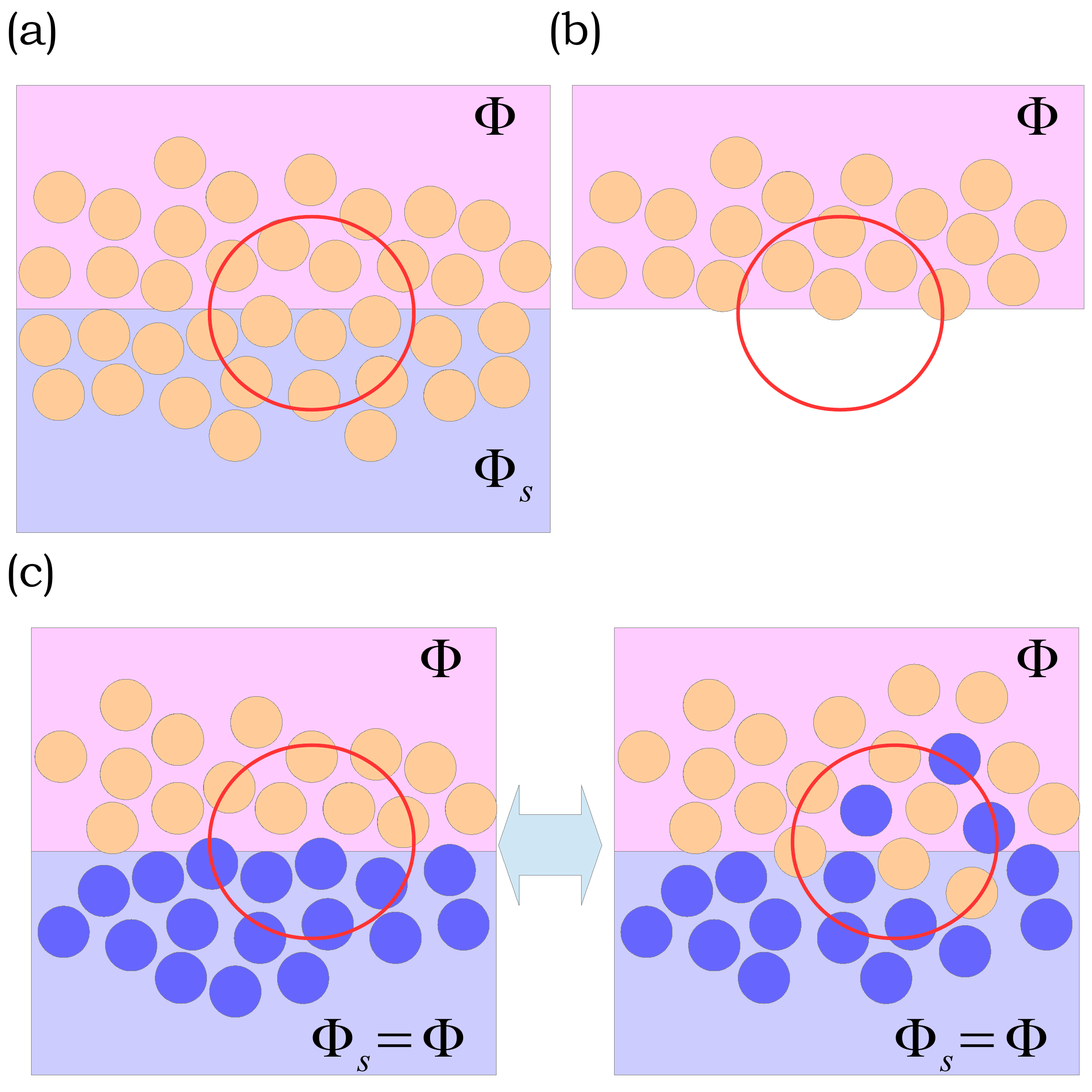}
\caption{\label{fig:3}(Color online) Illustration of different interfaces: (a) macroscopic bilayer with one sharp interface but (in general) different packing fractions in the two thick films, (b) free-standing or vapor interface thick film, and (c) pinned particle rough surface film which may or may not have the same packing fraction as the overlayer fluid. The image on the right indicates the technical approximation employed to map the real system to a first layer description based on an isotropic random pinned particle system \cite{66}.}
\end{figure}

The nature of the interface or substrate enters solely via the "surface" component of the caging dynamic free energy in Eq. (\ref{eq:19}). Per Eq(\ref{eq:18}), the modification of caging constraints at the surface always decreases at larger distances from the interface and bulk behavior is recovered deep enough into the thick film. We introduce 6 models for the "surface" component of the caging dynamic free energy that mimic to varying degrees of realism specific physical systems of experimental and simulation interest, as sketched in Fig. \ref{fig:3}. In each case there is a sharp interface between the liquid (top) and substrate (bottom) which are of macroscopic extent. Here we consider only physical systems where the dynamical structure of the substrate is a priori specified, i.e., the substrate sets boundary conditions and serves as an "external force field" felt by the liquid. We envision such models as directly relevant to simulations that employ pinned particle substrates, a film with a vapor interface, and as a simple model for amorphous or crystalline substrates (e.g., silica, silicon, gold) that are employed at temperatures far below their melting or glass transition temperature. Of course the latter can interact with the liquid via attractive interactions, and variable surface corrugation or roughness can play a role, surface effects not considered here.

\subsubsection{General Bilayer System}
This is the most general system considered here and is depicted in Fig. \ref{fig:3}a--two macroscopic layers (bilayer) with one sharp interface. The film and substrate are the same type of system (spheres of equal sizes) but, in general, can be at different volume fractions. In the first layer the cage center lies at the bilayer interface $(z=0)$. The dynamic free energy experienced by a liquid particle in this first layer is given by Eq. (\ref{eq:15}). This model does not address polymeric bilayers composed of two distinct glass-forming materials \cite{37,38,39} since in that case the bottom layer modifies the dynamics of the top layer and vice-versa. The bilayer system can be analyzed with our approach and will be studied in a future publication.

\subsubsection{Vapor interface}
Per Figure \ref{fig:3}b, since there are no particles in the vapor layer, one has:
\begin{eqnarray}
\Phi_s = 0, \quad F_{caging}^{surface}(r) = 0.
\label{eq:21}
\end{eqnarray}

\subsubsection{Rough pinned substrate}
Here the substrate is composed of literally pinned particles identical in every other way to the liquid particles. This is the simplest example of a rough solid substrate. It has been extensively studied in simulations and allows one to focus entirely on interface-induced changes of liquid dynamics. This system is of course anisotropic which renders the problem extremely complex. However, as discussed in Section III, the dynamic free energy of NLE theory is formulated at the cage scale based on an approximate angular averaging procedure. We implement this idea per the schematic of Fig.\ref{fig:3}c. At the interface, the half of a cage that are pinned particles are modeled as distributed randomly in a full spherical cage with the mobile particles. This zeroth order simplification assumes the most important consideration is the fraction of neighbors in a cage that are pinned versus mobile, and not their precise spatial arrangement. 

Given the above simplification, we employ our previously developed NLE theory for bulk randomly pinned particle fluids under neutral confinement conditions \cite{66} for a fraction of pinned particles in a cage equal to $\alpha = 0.5$.The dynamic caging free energy of this system is \cite{66}:
\begin{widetext}
\begin{eqnarray}
F_{caging}^{rough pinned}(\alpha,r) &=& -\int\frac{d\mathbf{q}}{(2\pi)^3}\left[ \frac{C(q)S_{12}(q)e^{-q^2r^2/6}}{\rho(1-\alpha)\left[1- \rho(1-\alpha)C(q)\right]}+
\frac{\rho(1-\alpha)C(q)^2e^{-q^2r^2\left[2-\rho(1-\alpha)C(q) \right]/6}}{\left[1- \rho(1-\alpha)C(q)\right]\left[2- \rho(1-\alpha)C(q)\right]} \right]_{\alpha=0.5} \nonumber\\
&=& \frac{1}{2}F_{caging}^{bulk}(r) + \frac{1}{2}F_{caging}^{pinned surface}(r),
\label{eq:22}
\end{eqnarray}
\end{widetext}
where $S_{12}(q)$ is the cross collective static structure factor between pinned and mobile hard spheres as discussed in Ref. \cite{66}. The second equality defines the rough pinned surface dynamic free energy:
\begin{eqnarray}
F_{caging}^{pinned surface}(r) = 2F_{caging}^{rough pinned}(0.5,r) -F_{caging}^{bulk}(r).
\label{eq:23}
\end{eqnarray}

\subsubsection{Rough vibrating pinned substrate}
A simple variant of model 3) allows the randomly pinned particles to harmonically vibrate via a small prescribed localization length $r_{L,s}$. This model is relevant to recent simulation studies of Simmons and co-workers \cite{29} that examined the influence of substrate mechanical stiffness or Debye-Waller factor on film dynamics. The parameter $r_{L,s}$ enters the theory via the first contribution on the right hand side of Eq. (\ref{eq:22}) which is modified by introducing the appropriate collective Debye-Waller factor of the vibrating pinned particles:

\begin{widetext}
\begin{eqnarray}
F_{caging}^{rough vibrating}(\alpha,r) &=& -\int\frac{d\mathbf{q}}{(2\pi)^3}\left[ \frac{C(q)S_{12}(q)e^{-q^2r^2/6}e^{-q^2r_{L,s}^2/6S(q)}}{\rho(1-\alpha)\left[1- \rho(1-\alpha)C(q)\right]}+
\frac{\rho(1-\alpha)C(q)^2e^{-q^2r^2\left[2-\rho(1-\alpha)C(q) \right]/6}}{\left[1- \rho(1-\alpha)C(q)\right]\left[2- \rho(1-\alpha)C(q)\right]} \right]_{\alpha=0.5} \nonumber\\
&=& \frac{1}{2}F_{caging}^{bulk}(r) + \frac{1}{2}F_{caging}^{vibrating surface}(r).
\label{eq:24}
\end{eqnarray}
\end{widetext}

\subsubsection{Smooth Rigid Wall}
For decades simulations have studied model supercooled liquids confined by a smooth hard wall (no corrugation, no attraction) which have no transverse wall-fluid forces. They find the locally anisotropic liquid dynamics is modified in a qualitatively different manner than for rough particle-based walls -- motion speeds up parallel to the wall and also in an angularly average manner relative to the bulk versus slowing down near rough walls of pinned particles \cite{13,58,67,68,69}. The hard smooth wall system can be viewed as simply a toy model, but it also may be crudely relevant to two classes of experimental systems as we briefly discuss.

Some hard substrates (dense crystalline or amorphous solids) are composed of atoms (size $b$) that are much smaller than the size of the molecules or polymer segments that constitute the fluid film, i.e., $b \ll d$. The substrate-fluid potential energy is thus of a corrugated form in the transverse direction which implies an oscillating (about zero) spatial variation of the wall-fluid forces on a length scale small compared to the fluid particles. If true, then for a nonadsorbing atomic substrate the transverse forces could average out to zero (at zeroth order) on the longer length and time scales relevant for the structural relaxation process of the larger fluid particles -- an effectively smooth wall. On the other hand, for short time and length scale dynamics (e.g., transient fluid particle localization near the substrate and "rattling dynamics") this picture will surely be less accurate and may not apply. 

Another experimental system perhaps related to the smooth wall model is a liquid substrate (e.g., glycerol \cite{40}) that is immiscible with the fluid film. It behaves in a thermodynamically hard manner corresponding to exerting repulsive forces on the fluid particles perpendicular to the interface. If the liquid substrate is of low enough viscosity such that its structural relaxation time is very small compared to the alpha time of the supercooled fluid, then the substrate particles are effectively ergodic from the perspective of the film particles. Within NLE theory this suggests an effective in-plane dynamical interfacial smoothness could apply on the alpha relaxation time scale of the film particles, and hence liquid substrates might be crudely viewed (to zeroth order) as a "smooth wall". Again such a viewpoint may not apply to the shorter time and length scale rattling dynamics of the fluid particles. 

In any case, our interest in this article is not the above two experimental systems, but solely to study the smooth wall surface as a limiting model where all wall-fluid forces parallel to the interface vanish. Compared to the rough pinned particle system, caging constraints exerted by the substrate on fluid particles in two spatial directions are absent. We crudely mimic this situation in an average manner by reducing the caging component of the dynamic free energy at the interface by a factor of 3. Thus, a smooth wall is modeled by dividing the pinned particle dynamic caging free energy by a factor of 3 corresponding to using in Eq.(\ref{eq:19}):
\begin{eqnarray}
F_{caging}^{smooth wall}(r) &=& \frac{1}{3}F_{caging}^{pinned surface}(r) \nonumber\\
&=&\frac{1}{3}\left[2F_{caging}^{rough pinned}(0.5,r) - F_{caging}^{bulk}(r)\right]\nonumber\\
\label{eq:25}
\end{eqnarray}

\subsubsection{Attractive Rough Walls}
We consider a variant of rough substrate models 3) and 4) where there is an attractive interaction between the mobile liquid and the immobilized substrate particles. Treating this fully is difficult given the high degree of nonuniversality of surface-fluid interactions, substrate structure, and the often presence of an explicit attractive force between the substrate and fluid particles. However, prior theoretical and simulation studies have found that a rather generic consequence of such an attraction is fluid densification near the wall, and typically only in the first layer \cite{13,67,70}. We consider a model that is a crude mimic solely of this effect by assigning a packing fraction in the first liquid layer that is higher than in other layers where it takes on the bulk value \cite{49}:
\begin{eqnarray}
\Phi_1 &=&\lambda\Phi_{bulk}, \qquad \lambda > 1 \nonumber\\
\Phi_j &=&\Phi_{bulk}, \qquad j \geq 2.
\label{eq:26}
\end{eqnarray}
Density enhancements are chemistry specific, but can be as large as 10-15 $\%$. As a specific example, for glycerol in contact with a silica surface, a recent computational study found \cite{70} a first layer enhancement of ~1.038. Alternatively, if the surface is weakly dewetting, the fluid density could be reduced, $\lambda < 1$. Explicit attractive forces are not taken into account dynamically, but they would serve to further slow down the mobile liquid particles near the surface. Treating the latter may require modifying the dynamic force vertex using the "projected dynamics theory" approach \cite{71}.

In subsequent sections we present representative numerical results for the key dynamical quantities of NLE theory for models 2), 3), 4), 5) and 6). A full analysis using ECNLE theory and the treatment of how various interfaces modify the collective elastic aspect of the alpha process in films is beyond the scope of this initial work and will be addressed in future publications.

\section{Short Length Scale Results: Dynamic Localization Length and Shear Modulus}
We first numerically apply the theory to study the most spatially local questions of the dynamic localization length for a thick film and the elastic shear modulus and its spatial gradient in a thin film.
\subsection{Dynamic Localization Length: Vapor vs Pinned Rough Solid Interfaces}
\begin{figure}[htp]
\includegraphics[width=8cm]{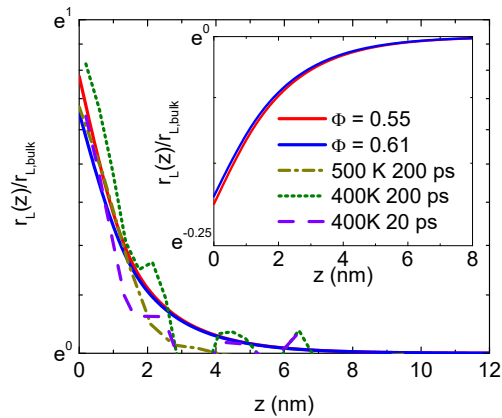}
\caption{\label{fig:4}(Color online) Spatial gradient of the dynamic localization length (normalized to its bulk value) for vapor interface (main frame) and pinned rough solid (inset) thick films. Results are shown in terms of an absolute length scale relevant to polystyrene ($d \sim 1.2$ nm) for $\Phi = 0.55$ and 0.61. We find that the normalized gradient localization lengths for the vapor interface and pinned rough solid films at $\Phi = 0.55$ are well described by the exponential fits $r_L(z)/r_{L,bulk}=1+1.28469e^{-z/0.832d}$ and $1-0.18286e^{-z/1.63d}$, respectively. The dotted, dashed and dot-dashed curves are simulation results of \cite{31} based on using MSD data evaluated at the time scales and temperatures indicated in the legend of the inset. The inset shows the same theoretical results plotted in a natural log-linear format.}
\end{figure}

Figure \ref{fig:4} shows the spatial variation of the dynamic localization length, $r_L$, normalized to its bulk value for two very different values of volume fraction for vapor interface (main frame) and rough pinned solid interface (inset) thick films. For both systems, this relative dependence depends very weakly on volume fraction, reflecting the near "factorization" property of the NLE dynamic free energy discussed in sections II and III. As expected, the localization length is larger (smaller) near the vapor (solid) surface. Moreover (see the inset and figure caption), we find that it decays to the bulk value in an exponential manner with an essentially volume-fraction-independent characteristic length scale of $\sim 0.83$ and 1.7 particle diameters for vapor and solid surfaces, respectively. These results are akin, at zeroth order, with Eq.(\ref{eq:20}) that suggests a decay length of $\sim d/ln(2) \sim 1.45d$. The shorter penetration length for the solid surface compared to the "softer" vapor interface is interesting, especially since the amplitude of the surface perturbation (deviation of $r_L(0)/r_{L,bulk}$) from unit is larger for the vapor film. This reveals a nontrivial discrimination by the theory between the surface amplitude versus penetration depth aspects of soft and hard interfaces. Moreover, the predicted trend for the localization length appears to be in qualitative accord with the experimental finding that the spatial range of $T_g$ perturbations near soft interfaces are greater than for hard interfaces \cite{38,39}.  

Our results for the vapor interface are also compared (with no fitting) to the recent free-standing film simulations of an atomistic polystyrene liquid model of Zhou and Milner \cite{31}. Although there is some ambiguity associated with the extraction of a transient dynamic localization length via the intermediate time segmental mean square displacement (MSD) in simulation \cite{31}, there is broad consistency between the data and theory including the relative insensitivity to thermodynamic state, the exponential decay, and the amplitude of the change at the surface. Note that the MSD normalized by its bulk value is nearly independent of temperature over the range $400K-500K$. This agrees well with our predicted weak density dependence of $r_L(z)/r_{L,bulk}$. The normalized gradient of the simulation also seems to be insensitive to the analyzed MSD time.

Figure A1 shows our corresponding predictions using the inhomogeneous film NMCT of section IIIA. One sees very good agreement with the dynamic free energy based analogs for the vapor interface film, but a significantly shorter range gradient for the pinned solid surface system. Another important point is the comparsion to the prior NLE-based theory of MS \cite{47,48,49}. Figure \ref{fig:11} shows the localization length enhancement is of very similar magnitude near the surface, but decays much more quickly to the bulk value at a distance of $\sim 1.3-1.5d$, as expected. 

\subsection{Smooth Wall and Vibrating Rough Particle Interfaces}
\begin{figure}[htp]
\includegraphics[width=8cm]{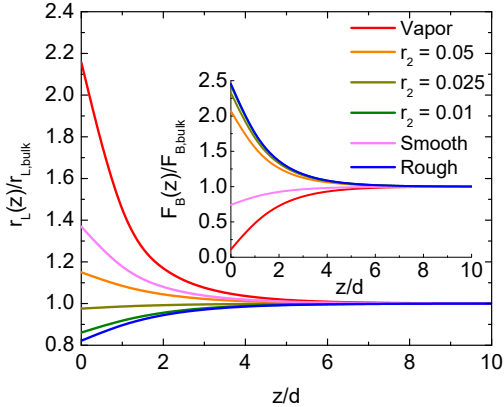}
\caption{\label{fig:5}(Color online) Normalized (to the bulk value) localization length gradient at $\Phi = 0.57$ for 4 types of indicated surfaces: vapor, and vibrated pinned and literally pinned solid treated in two ways to mimic a rough and smooth hard surface. The vibrating pinned rough substrate results are shown for the 3 indicated small values of vibrational amplitude (surface localization length). Inset: The analogous calculations for the normalized local cage barrier, $F_B(z)/F_{B,bulk}$ for smooth and vibrating surfaces with $r_{L,S}/d =0.05, 0.025, 0.01$, which we find are well fit by the exponential forms $1 - 0.258e^{-z/1.523d}$, $1 + 1.067e^{-z/1.333d}$, $1 +1.324e^{-z/1.351d}$, and $1 +1.4298e^{-z/1.355d}$, respectively.}
\end{figure}

Figure \ref{fig:5} shows representative calculations of the normalized localization length gradient for the smooth wall and vibrating particle rough wall (for 3 values of surface particle localization lengths) models, and contrasts them with the vapor and pinned particle results of the previous section. Interestingly, the smooth hard wall system exhibits enhancements of the localization length, and hence behaves more akin to a vapor interface than a rough pinned particle substrate for this property. The vibrating particle rough wall systems evolve from suppression of the localization length for small vibrational amplitude ($r_{Ls}=0.01d$), to weak enhancement for large vibrational amplitude ($r_{Ls}=0.05d$). We find that all the systems studied show a good exponential decay profile (see the caption for fit functions), with a characteristic length scale of $\sim 1-2d$. 

\subsection{Elastic Modulus}
\begin{figure}[htp]
\includegraphics[width=8cm]{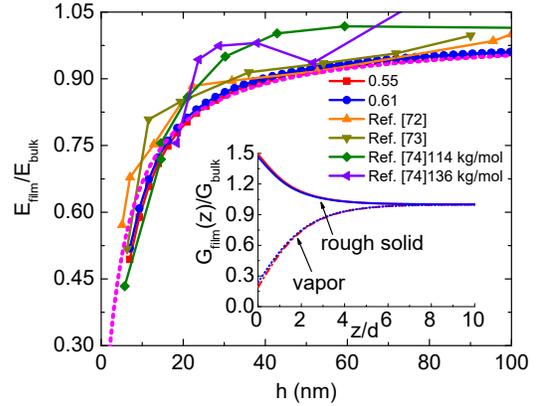}
\caption{\label{fig:6}(Color online) Main frame: Normalized to the bulk film-averaged elastic moduli of a free-standing vapor interface film at volume fractions $\Phi = 0.55$ (red points) and $\Phi = 0.61$ (blue points) as a function of film thickness in nm ($d=1.2$ nm per PS). Orange and dark yellow curves correspond to MD simulation data for PMMA \cite{73} and 33DDS+DGEBA \cite{72} polymer films, respectively. Dark green and purple curves are experimental data \cite{74} for a PS polymer film floated on water having weight-average molecular weights of 114 kg/mol and 136 kg/mol, respectively. The dotted red curve is the empirical analytic function $E_{film}/E_{bulk} = 1/\left(1+\delta^E/h \right)$ with $\delta^E = 4.2$ nm. Inset: theoretical normalized dynamic shear modulus gradient of both the vapor interface and pinned particle solid surface films for $\Phi = 0.55$ (red points) and $\Phi = 0.61$ (blue points).}
\end{figure}

We employ Eq.(\ref{eq:4}) with a $z$-dependent localization length to calculate the glassy elastic modulus gradient. Results are shown in the inset of Fig. \ref{fig:6} in the format of shear modulus at location z divided by its bulk analog for two volume fractions. The value of the latter does not matter in a practical sense in the normalized format. Visually, the glassy modulus gradient extends 4-5 particle diameters into the film. The modulus softens at the vapor surface by a factor of $\sim 3$, while at the pinned rough surface there is hardening by $\sim 50 \%$.  

The main frame of Fig. \ref{fig:6} shows calculations of the film-averaged elastic modulus normalized by its bulk value. Given we have not formulated an explicit theory for the 2-interface thin film, we have crudely assumed two independent gradients emanating from each vapor surface which do not interfere. Moreover, the film-averaged modulus has been computed using an arithmetic average per a "series" model. Given the importance of the question of how spatial gradients of dynamic properties are properly weighted in thin films \cite{18,29,48,49}, a "parallel" averaging mechanical model may be more appropriate, but studying this issue is beyond the scope of this article. 

The calculations have been done at two hard sphere volume fractions, with effectively the same results found. The theoretical spatial gradients for hard spheres are naturally represented in terms of $z/d$, where $d$ is the effective particle diameter. Conversion of the x-axis to real units allows comparison with experimental data on polystyrene films \cite{54}; we use the known value of Kuhn segment diameter, $d \sim 1.2$ nm \cite{45,46}. The theory predicts the modulus softens by a factor of $\sim 2$ for a $\sim 10$ nm thick film, and bulk behavior is recovered only for films approaching 100 nm thick. 

Various experimental and simulation data sets are also shown in Fig. \ref{fig:6}. Although there is some disagreement among experimentalists \cite{36,53}, all simulations and the large majority of experimental studies find a vapor interface induces a softening of the elastic modulus near the surface. The simulations of \cite{72} and \cite{73} employed coarse-grained molecular dynamics (CGMD) to compute the size-dependent Young’s modulus of the polymer diglycidyl ether Bisphenol with 3.3’-diaminodiphenyl sulfone (DGEBA/33DDS) and polymethylmethacrylate (PMMA) free-standing films, respectively. Since our gradients of normalized localization length agree with the simulations \cite{31} (see Fig. \ref{fig:4}), and given Eq.(\ref{eq:4}), we expect good agreement between theory and simulation for the spatially-dependent Young's modulus normalized by its bulk value, $E(z)/E_{bulk}$. This expectation is verified in Fig. \ref{fig:6}. 

The (rather noisy) experimental data shown is for polystyrene thin films \cite{74} of thicknesses that range from from 7 nm to 220 nm. The films were deposited on water to avoid gravitational deformation. Measurements of stress-strain response until the polymer film breaks in a brittle manner are used to extract Young’s modulus. The averaged experimental moduli ratio depend to some extent on sample width. These real world complications introduce some uncertainty in comparing to our theoretical calculations based on linear response and two vapor interfaces to the experimental data. Nevertheless, there is rough consistency between theory and experiment for the magnitude of modulus changes and variation with film thickness. We note that the bulk value of the modulus is experimentally recovered at a smaller film thickness than in our calculations, but nearly quantitative agreement is found for thicknesses of 20 nm and smaller. 

Given we find that the theoretical localization length is well described by an exponential decay function, it is not surprising that we find the modulus gradients of Fig. \ref{fig:6} also follow an exponential form to a high degree of accuracy (not shown). However, as known from prior experimental and simulation studies, other functional forms can also fit well the data. As an example of this point, the dotted curve in the main frame shows that our film-averaged normalized elastic modulus results can also be well fit using a popular empirical function in the literature, $1/(1+\delta^E/h)$.

\section{Jump Distance and Local Cage Barrier:  Vapor, Pinned Rough and Smooth Solid Interfaces}

\begin{figure}[htp]
\includegraphics[width=8cm]{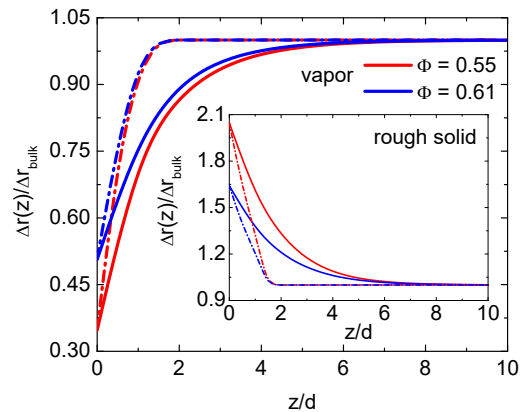}
\caption{\label{fig:7}(Color online) The particle jump distance normalized to its bulk value for vapor interface (main frame) and pinned particle rough surface (inset) thick films as a function of distance from interface for $\Phi = 0.55$ and $0.61$. The solid curves correspond to calculations using the new dynamic cage constraint transfer idea of this article, and the dashed-dot curves are based on the prior theory \cite{47,48,49} which did not include this effect. We find that the ratio $\Delta r(z)/\Delta r_{bulk}$ for the vapor interface films is well fit by $1-0.64563e^{-z/1.15d}$ and $1-0.49249e^{-z/1.25d}$ for $\Phi = 0.55$ and $0.61$, respectively, and the pinned particle rough surface film results are well fit by $1+1.053e^{-z/1.644d}$ and $1+0.642e^{-z/1.702d}$ for $\Phi = 0.55$ and $0.61$, respectively.} 
\end{figure}

A crucial additional dynamical property needed to quantify the elastic barrier in ECNLE theory is the effective jump distance of Eqs. (\ref{eq:5}) and (\ref{eq:6}). To predict the alpha time gradient also requires knowledge of the local cage barrier gradient, $F_B(z)$. In this section, we use the theory to study these two dynamical properties in films with vapor, pinned rough, and smooth hard wall surfaces.

\begin{figure}[htp]
\includegraphics[width=8cm]{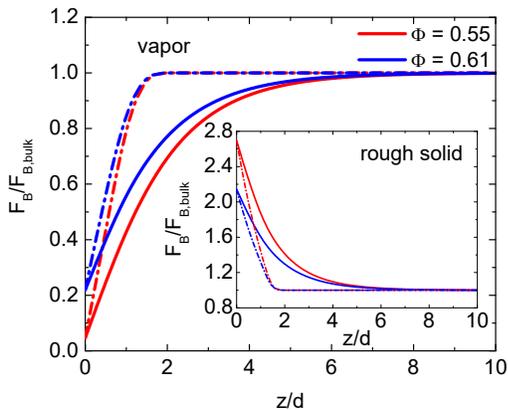}
\caption{\label{fig:8}(Color online) The spatial gradient of the local cage barrier height normalized by its bulk value for vapor interface (main frame) and pinned particle rough surface (inset) thick films as a function of distance from the interface at $\Phi = 0.55$ and $0.61$. The solid curves correspond to the new theory results and the dashed-dot curves are those of the prior theory \cite{47,48,49} that ignored the longer range cage constraint transfer effect. We find that the ratio $F_B(z)/F_{B,bulk}$ of vapor interface films are well fit by $1-0.9647e^{-z/1.66d}$ and $1-0.784e^{-z/1.539d}$ for $\Phi = 0.55$ and $0.61$, respectively, and the pinned particle rough surface results are well fit by $1+1.691e^{-z/1.325}$ and $1+1.14e^{-z/1.395d}$ for $\Phi = 0.55$ and $0.61$, respectively.} 
\end{figure}

The main frames of Fig. \ref{fig:7} and \ref{fig:8} show results for the above two quantities at two volume fractions for the vapor and pinned solid interface models. Also shown for comparison are the analogous results for the vapor interface based on the simpler MS \cite{47,48,49} approach. For a vapor interface, the jump distance (Fig. \ref{fig:7}) and local barrier (Fig. \ref{fig:8}) are strongly reduced at the surface, and more so at lower packing fraction. The gradients visibly decay on a length scale of $\sim 5d$. We find they are all well fit by an exponential function (see figure captions) with decay lengths in the range of $\sim 1-2d$. The latter depend relatively weakly on property and interface, and almost not at all on volume fraction, trends which can be understood from the general nature of the theory discussed in section IIIB. Although the direction of changes of these properties at the surface are the same as in the prior approach \cite{47,48,49}, incorporation of longer range mobility transfer physics leads to a much slower spatial decay with a different functional form. 

The insets of Fig. \ref{fig:7} and \ref{fig:8} show the analogous results for the rough pinned solid surface. The qualitative trends of the gradients, compared to each other and to the prior more local approach \cite{47,48,49}, are the same as found for the vapor surface, although the exponential decays lengths are non-trivially larger. On the other hand, the relative enhancement of the two properties at the surface is a factor of $\sim 2-2.5$ smaller than for a vapor interface. These are the same relative trends as found for the dynamic localization length in Fig. \ref{fig:4}. In all cases, changes in the jump distance near the surface are large which will have big consequences in the prediction of the alpha time gradient since this length scale enters as the 4th power in determining the elastic barrier per Eqs(\ref{eq:5}) and (\ref{eq:6}).

Figures \ref{eq:4}, \ref{eq:7} and \ref{eq:8} compare results for the localization length, jump distance and local barrier height for the vapor and rough solid surfaces at two different volume fractions. We have studied these questions over a wide range of volume fractions, and the general trends found are consistent with the representative results in the aforementioned figures. For example, as volume fraction grows from 0.55 to 0.62, for the vapor surface the ratio of $r_L$, $\Delta r$, and $F_B$ at the surface to their corresponding bulk values vary monotonically over the range from $\sim 2.3-2$, $\sim 0.35-0.53$ and $\sim 0.05-0.3$, respectively; the corresponding values for the pinned surface are $\sim 0.8-0.85$, $\sim 2.1-1.65$ and $\sim 2.7-2.1$. One sees that for all properties there is a stronger volume fraction dependence for the vapor interface system.

\begin{figure}[htp]
\includegraphics[width=8cm]{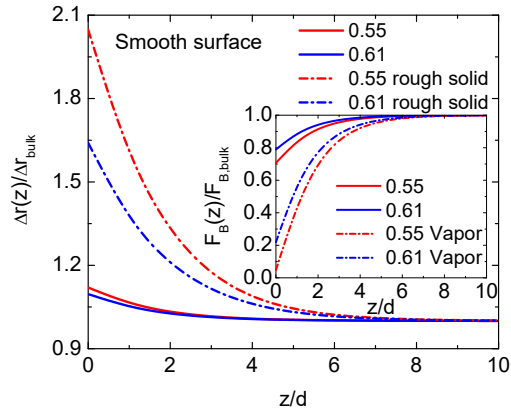}
\caption{\label{fig:9}(Color online) The normalized (by its bulk value) jump distance calculated for pinned particle rough surface (dot-dashed curves) and smooth hard wall surface (solid curves) thick films as a function of distance from interface  for $\Phi = 0.55$ and $0.61$. We find that the ratio $\Delta r(z)/\Delta r_{bulk}$ for the two smooth hard wall films are, to leading order, both well fit by $1+0.12e^{-z/1.472d}$ for $\Phi = 0.55$. The inset shows the analogous calculations for the normalized local barriers in vapor interface (solid curves) and smooth hard wall (dash-dot curves) films.} 
\end{figure}

We now consider smooth hard walls. The localization length calculations of Fig. \ref{fig:5} suggest this system behaves in a manner intermediate between a vapor surface and a rough solid surface. Figure \ref{fig:9} shows calculations of the normalized jump distance (main frame) and local barrier (inset), and contrasts the results with the vapor and two different solid interface analogs. The latter two systems exhibit a large suppression and enhancement of the jump distance, respectively. The smooth surface shows only a very weak enhancement of this quantity, but the form of the spatial decay is again exponential and of a range similar to that of the other two systems (see caption). While the normalized local barrier in the inset of Fig. \ref{fig:9} qualitatively behaves as if the smooth surface was more like a vapor interface, its suppression is quantitatively much weaker. Given the smooth surface shows an enhanced jump distance compared to the bulk (which would increase the collective elastic barrier in ECNLE theory) but shows a smaller local cage barrier, how the alpha time mobility gradient will change is subtle and unclear. But since all changes for the smooth surface relative to the bulk are rather small, one expects the mobility modifications for this system will be modest. In the experimental polymer film community, such a situation has been inferred, for example, for polystyrene films supported on substrates such as silicon and silica \cite{1,2,3}. The phrase "neutral substrate" is typically invoked to indicate a hard surface that has little effect on the dynamics or $T_g$ of the film.

The inset of Fig. \ref{eq:5} shows results for the local cage barrier at a representative volume fraction of 0.57 for a smooth hard wall compared to the other systems; in the normalized barrier format shown the results are nearly independent of packing fraction although small variations are typically found in the high packing fraction regime (0.55-0.61). Interestingly, the smooth wall system now shows a suppression of the local barrier, albeit rather weak. In conjunction with the smooth wall results in Fig. \ref{fig:8}, this again buttresses the view that the smooth wall model may be relevant to nearly atomically smooth hard surfaces (e.g., silica, silicon) where molecule-surface or polymer-surface adhesion is weak---a "neutral hard surface". One also sees that allowing pinned particles to vibrate modestly reduces the local cage barrier, but the degree of change relative to the bulk is smaller than for the localization length. These trends seem physically sensible given the barrier is determined by motion on a length scale far beyond a vibrational amplitude. But for all systems, the spatial range of the local barrier gradients are essentially the same, and the same as the other key features of the dynamic free energy. Bulk behavior is recovered in a practical sense at $\sim 4-6$ particle diameters into the film.

\begin{figure}[htp]
\includegraphics[width=8cm]{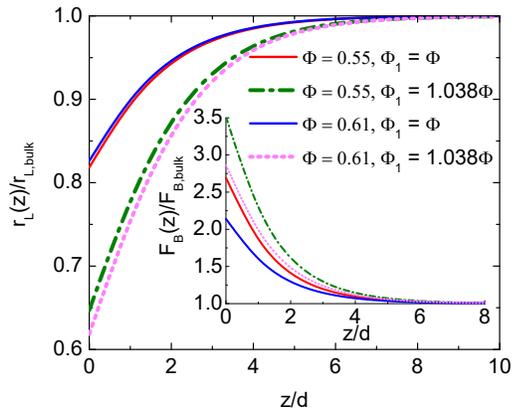}
\caption{\label{fig:10}(Color online) Main frame: Normalized (to the bulk value) dynamic localization length gradient for pinned particle rough surface films as a function of distance from interface at $\Phi = 0.55$ and $0.61$ with (dashed and dotted curves) and without (solid curves) densification in the first layer. Inset: The analogous calculations for the normalized local cage barrier. The first layer density enhancement factor is 1.038. The barriers in the bulk are 4.7 and 12.9 in thermal energy units.} 
\end{figure}

Finally, Figure \ref{fig:10} presents one example of how surface-induced densification of the first liquid layer affects the dynamic localization length and local cage barrier. The chosen value of density enhancement of 3.8 $\%$ is motivated by a recent computational study of liquid glycerol exposed to a silica surface \cite{66}.  Calculations are shown for two values of volume fraction, 0.55 and 0.61, where the bulk local barrier in thermal energy units is 4.7 and 12.9, respectively. The corresponding results if there is no first layer densification are shown for comparison.

The main frame of Fig. \ref{fig:10} shows that such a modest densification results in a major enhancement of particle localization near the hard surface. However, the changes relative to the bulk are almost the same at the two different volume fractions studied with and without densification. Moreover, the length scale for visually recovering bulk behavior is almost the same for all calculations, $\sim$ 6-7 particle diameters. Such densification is expected to result in an increase of any practical measure of the gradient width. The inset shows the analogous calculations for the local barrier. The absolute and relative trends are qualitatively the same as found for the localization length in the sense that densification results in slower dynamics (higher barrier). Given knowledge of the bulk barrier height, this degree of enhancement translates to a barrier in the densified first layer that is larger by roughly 3.8 $k_BT$ (10.3 $k_BT$) for the lower (higher) volume fraction system. Since the alpha relaxation time scales as the exponential of the barrier, even only taking into account this change of the local cage barrier would result in an increase of the alpha time by a factor of $\sim 45$ or $\sim 30,000$, respectively. 

\section{Discussion}
We have constructed a new particle-level microscopic theory for how dynamic caging constraints at a surface or interface are modified and spatially transferred in a layer-by-layer manner into the film interior in the context of the dynamic free energy concept of the force-based NLE theory. The basic idea is to reduce the resolution of the cage level description to acknowledge different dynamical constraints in the upper and lower halves of a cage. The effective dynamic free energy at any mean location (cage center) then involves contributions from two adjacent regions where confining forces are not the same. The $z$-dependence of the caging component of the dynamic free energy varies essentially exponentially as a function of distance from the interface, with a universal decay length of modest size and weak sensitivity to thermodynamic state. Such a variation imparts a roughly exponential variation of all key features of the dynamic free energy required to treat dynamical gradients of the localization length, jump distance, cage barrier, and alpha time. As an important consequence we expect that, to leading order, a double exponential form of the alpha time spatial gradient is predicted.

Diverse systems were considered where the surface was a vapor, a rough pinned particle solid, a vibrating (softened) pinned particle solid, a smooth hard wall, and a solid substrate which densifies the first layer of the liquid. The fundamental manner that they enter the theory at the level of the dynamic free energy is the same, with the crucial difference arising solely from the first layer where the non-universal dynamical constraints can be weaken, softened, or hardly changed depending on the interface. However, both the amplitude of the modification and its quantitative spatial range of penetration into the film varies with interface type, although the penetration depth is weakly dependent on density or temperature. Numerical calculations for the hard sphere fluid established the spatial dependence and volume fraction sensitivity of the changes of key dynamical properties for 5 different models. No adjustable comparison of the theoretical predictions for the dynamic localization length and glassy modulus against simulation and experiment for systems with vapor surface(s) reveal good agreement.

Future work will fully integrate the new advance reported in this article with all aspects of ECNLE theory for films with vapor and solid interface films including the collective elasticity contribution. This will allow us to make quantitative predictions for quantities such as the alpha relaxation time gradient, dynamic decoupling phenomena, $T_g$ gradient, and film-averaged properties for both model systems and experimental materials with diverse interfaces and chemical nature of the building blocks (colloids, molecules, polymers) Key open questions such as the near double exponential variation of the alpha time gradient, how the amplitude of the alpha time changes at the surface, the length scale of the dynamic gradient, how the apparent decoupling exponent \cite{57} precisely varies with location in a film, and the role of solid substrate elasticity or Debye-Waller factor \cite{29} will be addressed in detail for vapor, pinned particle solid and other interfaces. The impact of the now longer range nature of surface-induced changes of dynamics emanating from the interface compared to the prior formulation of the ECNLE theory of thin films \cite{47,48,49,50} on how important the cutoff at the interface of the collective elastic component of the alpha process is will be re-visited. Finally, the basic new idea of the present paper is generalizable to different confined geometries (spherical droplets, cylindrical pores) and polymer or molecular bilayers.
\appendix
\counterwithin{figure}{section}
\section{NMCT formulation of Dynamic Localization Lengths in Films}

\begin{figure}[htp]
\includegraphics[width=8cm]{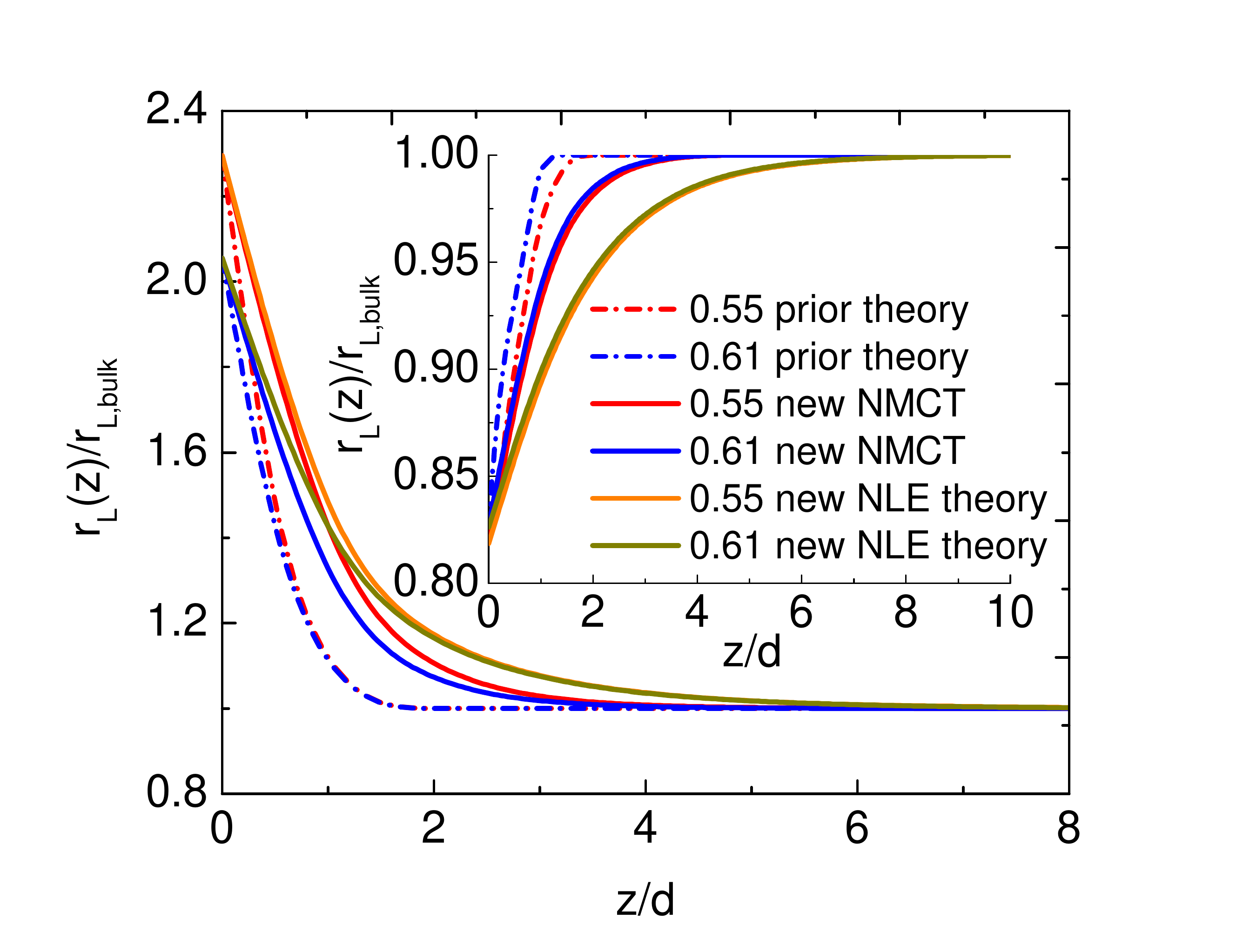}
\caption{\label{fig:11}(Color online)Normalized gradient of the dynamic localization length, $r_L(z)/r_{L,bulk}$, for vapor interface (main frame) and rough pinned particle surface (inset) thick films calculated using NLE theory with the dynamic free energy concept without (dashed-dot curves) and with (solid orange and green curves).  the new cage constraint transfer effect. The analogous latter results based on the ideal NMCT formulation of Eq. (\ref{eq:11}) are shown as the solid red and blue curves. } 
\end{figure}

The normalized dynamic localization length gradients are calculated using the NMCT formulation of Eqs.(\ref{eq:3}) and (\ref{eq:4}), and the results are compared to the layer-based NLE dynamic free energy formulation. Figure \ref{fig:11} shows representative results, and ones sees the two theories make very similar predictions for the vapor interface, but there are quantitative differences for the pinned particle surface model. All results can be described by an exponential function. Both calculations show an insensitivity of the normalized gradient to the fluid packing fraction. Also shown are the predictions of the prior formulation of MS \cite{47,48,49} for a vapor interface which assumed surface-nucleated reduction of the caging constraints is confined to a distance of only $r_{cage}\sim 1.5d$ from the interface. Obviously including the new physics developed in this work greatly extends the spatial modification of $r_L(z)/r_{L,bulk}$ relative to this prior formulation.	

\begin{acknowledgments}
This work was supported by DOE-BES under Grant No. DE-FG02-07ER46471 administered through the Frederick Seitz Materials Research Laboratory. We thank Professor David Simmons for stimulating and informative discussions and for sending us a preprint of Ref.\cite{57}. We thank Dr. Yuxing Zhou for providing the simulation data from Ref.\cite{31} and helpful discussions. 
\end{acknowledgments}

\end{document}